# Loklak - A Distributed Crawler and Data Harvester for Overcoming Rate Limits

Sudheesh Singanamalla[*]
Microsoft Research
t-sus@microsoft.com

Michael Peter Christen[†]
YaCy.net
mc@yacy.net

## ABSTRACT
Modern social networks have become sources for vast quantities of data. Having access to such big data can be very useful for various researchers and data scientists. In this paper we describe Loklak, an open source distributed peer to peer crawler and scraper for supporting such research on platforms like Twitter, Weibo and other social networks. Social networks such as Twitter and Weibo pose various limitations to the user on the rate at which one could freely collect such data for research. Our crawler enables researchers to continuously collect data while overcoming the barriers of authentication and rate limits imposed to provide a repository of open data as a service.

## 1. INTRODUCTION
Online social networks have changed the way people communicate across the world and have thus become very important for various research communities working in the fields of information extraction, social network analysis, natural language processing, behavioural analysis and market trend predictions while leveraging the nature of virality present in these social networks. Obtaining the data from such social networks using the official APIs require authentication and are rate limited thus making data collection a very complex problem. Researchers working in these areas tend to spend a lot of time collecting the required data and build systems for obtaining and storing this data for further analysis.

The rate limits imposed by such services affect the research work carried out by researchers and data analysts where significant amount of effort is put into gathering the data. Many studies on the usage of hashtags [2], mentions, user interactions [5] as well as specific content analysis have made an effort to crawl data independently for research purposes and a significant effort has been made to make the datasets openly available for the wider scientific community to promote further research. The studies have also shown that scarcity of such publicly available datasets in other languages like Arabic [3] also pose a problem to many researchers. Twitter's access to APIs is severely limited by the number of queries that can be made in a fixed interval of time thus making the data collection process a more time consuming process.

Twitter REST API exposes various rate limited services for accessing the public content of the user profiles and their messages. A rate limit window[1] denotes the time window in which the number of queries can be made and is used to control the traffic to the given endpoint. Twitter denotes the rate window to be fifteen minutes and allows a maximum of fifteen queries in that specific time window. Each of these API requests made to the REST endpoints need an authentication token obtained by registering an application with twitter and the number of tokens issued is limited to 100,000. The data APIs like **statuses/home_timeline** allow users to fetch up to 800 records in the fifteen minute time window and is severely rate limited. At the same time the passing of the user context/token retrieves only the tweets from the followers on the application owners account and is dependant on the number of followers the owner has.

Loklak is an open source[2] approach that leverages the power of the open source community and peer to peer architectures to openly share data and crawl twitter and other such micro blogging services. This tool allows researchers and open data communities to easily host the servers with one click on various easily available hosting platforms like Azure, Heroku, Scalingo, Bluemix or any other Docker based systems. The modular nature of the Loklak server allows it to be easily extensible to various other micro blogging websites while at the same time allowing flexibility in the way the REST responses are obtained thus making it an easy replacement for various applications using the official Twitter API. The peer to peer nature of the application helps to maintain clusters and host them in different cloud regions while allowing the maintainers to adapt any specific network architecture to run the crawlers without getting blocked out by Twitter or such micro blogging services for performing too many queries.

## 2. RELATED WORK
This section reviews prior work done in building crawlers and existing tools which were available to support research on twitter data in the recent years. The first large scale crawl of twitter was done by Kwak et al. [5] to study the network topology of Twitter and its power for information sharing. All of these crawls involved usage of the Twitter API with white listed machines to perform the crawl while using a self imposed data ingestion limit on the system. According to the recent guidelines Twitter no longer provides any whitelisted accounts.

There has been previous effort in building software tools that allow researchers to crawl twitter by themselves while using the existing APIs but this approach is severely limited by the query limits and window size imposed by Twitter, one such tool has been presented at TREC 2011 Microblog track[1][3]. The alternative is to take up paid services like Gnip which cost 2000 USD per month for enterprise users and 300 USD per month for a personal small sandbox environment.

---
[*]Also affiliated with Loklak and FOSSASIA
[†]Initiator of the YaCy and Loklak projects

[1]https://dev.twitter.com/rest/public/rate-limiting
[2]https://github.com/loklak/

One of the first data collection using crawlers without white lists was done by Java et Al [4] in 2007 from April 1 to May 30th 2007 which resulted in the collection of 1.3 million tweets from 76 thousand users. With the changes in the user interfaces and animations on Twitter this crawl would no longer work with the same efficiency with which it has worked before. TwitterEcho [1] proposed a distributed focused crawler to support open research with twitter data and talks about using web crawlers to collect the required information from such social networks.

Many other researchers have crawled twitter and used the data to investigate events from messages in Arabic with the EVEtar system, [3] digital epitemology, [4] improving influenza forecasting, [6] various visualization tools [7] and even in understanding human behaviour. [8]

From the research done so far we can conclude that there has been a lot of care taken in building systems that respect the limits set by Twitter and have all been done in controlled and experimental laboratory settings than at a production scale over the Internet. The crawls done have only been done for short durations of few months by Java et Al [4] and Wang [9] from January 3rd to 24th 2010 compared to the 2 years of crawling and indexing by the system described in this paper. The usage of distributed systems is required to collect large amounts of data. [1, 5, 9] As far as we know we are the first system that is capable of crawling and scraping information from twitter at a fast rate overcoming the rate limits and without authentication.

## 3. ARCHITECTURE

Loklak is a distributed peer to peer back end storage infrastructure for microblogging messages and uses elasticsearch as its main back end resource. It exposes similar APIs like the Twitter APIs and can be instructed to perform a crawl and provide the responses or fetch the responses from the peers or from its own back end storage. It leverages the advantages of distributed systems to improve the rate of crawling and performs data collection on such rate limited services using scraping thus removing the bottleneck of authentication. It is written in Java and runs on most platforms like Windows, Linux and Mac OS X along with single click to deploy procedures to various cloud hosting providers like Azure, Bluemix, Heroku and Docker cloud. The default back end storage to push the crawled tweets to is registered as the Loklak webserver which allows a crowd sourced twitter data collection and can be easily customized to push the data from the peers to a self hosted Loklak cluster instead of the pre-configured back end source.

Each of these deployments act as peers in the network and maintains local databases that store the results of the crawls together and periodically pushes the stored objects to the head peer, however it is also possible to run Loklak as a standalone server on a single machine without creating a network or designating a head peer. Such a system is much more prone to being blocked out from twitter over time if the request limit isn't set from the server manually as a part of its configuration. The server application is highly modular and is designed to easily allow the crawling and scraping of other similar websites like Weibo and Instagram.

[3]https://sites.google.com/site/microblogtrack/
[4]https://blog.twitter.com/2015/twitter-data-public-health

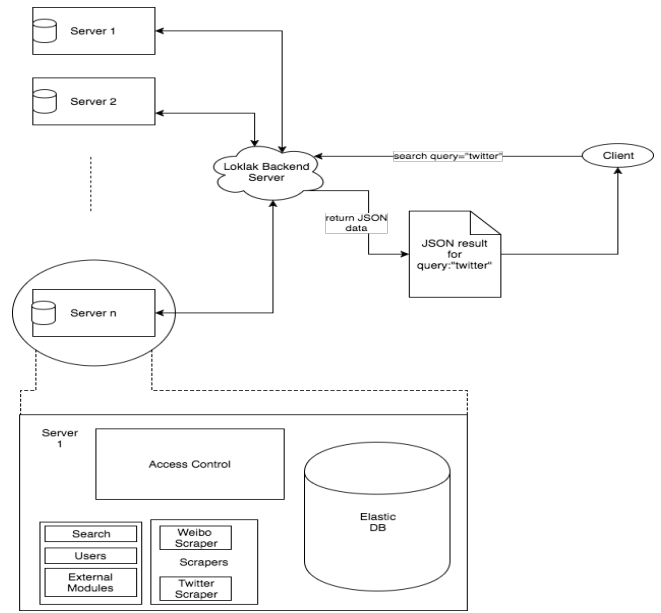

Figure 1: Client request for data

### 3.1 Clients & Instructors

Each of the Loklak servers hosted contain a client which makes the required requests and scrapes the response HTML into objects from the corresponding classes and attributes. All the clients crawl the twitter application by recursively queueing up each of the *mentions, hashtags* and *query* text up to a specific configurable crawl depth. The client in the network notifies the back end by making a join query to it. After a successful peer-join acknowledgement the back end allows the peer to post the data from its local database to the back end database thus crowd sourcing the information across the peer to peer network.

When a client makes a search request from a Loklak instance in the peer to peer network, the instance requested makes a request for the data within its local database as well as creates a *fetch* request to all the peers available in its network. The instance can then trigger multiple peers on the network to perform a crawl for the data and send the data back to the Loklak instance that the client has requested.

When a client sends a request for a specific search term it has the ability to request the instance to also perform a re-crawl for the given query. In Figure 1 the client has requested the main back end server. The connected peers can share the information using the peer interfaces within the server instances as explained in the next subsection. In this case, the client has requested tweets with the keyword *twitter* to a back end server, this server looks up for the data in its local database and also sends an asynchronous request to its peers with the same query behaving as a client. Any information from the peers is then pushed from the local databases of each of the peers to the database of the back end server after removing duplicate entries in the data obtained from the peers. The client has different ways to receive the data response for the query made across the network and can also instruct the servers and peers to perform crawls at a particular depth.

A client can issue multiple queries or instructions to the server instance from which it requests the data. The server broadcasts every query to its peers requesting for more information in an effort to update its local database. When the request is only for the data from the local database of the requested instance, the requested instance immediately responds with the data after broadcasting an *update* request without waiting for a response from its peers thus updating the database later as the response from the peers are obtained. When the client requests data including from it's peers, the response is obtained only after all peers respond or after a fixed time interval in case a peer fails to respond. Added to the querying ability the client has the ability to trigger a set of instructions to the server instance to asynchronously crawl and scrape the required information from the social network. As more server instances are geographically dispersed and form the peer to peer network more information can be collected about local micro blogging trends of different countries.

## 3.2 Server Instances

Each of the server instances contains multiple modules in its architecture. Every request is processed first by the Access Servlet. Depending on the type of client, the server provides a different level of access on the data and instruction control of the server. All localhost clients are given more access to the server and the ability to launch crawl jobs at greater recursive crawl depth than the peers or clients instructing the server instance.

A caretaker spins up concurrent threads that perform the role of asynchronously doing the peer to peer operations and transmissions. After determining the access level of the client, the server checks the request frequency to prevent a DDoS attack.

The client query is then processed by the REST interface, depending on the type of the client with localhost and privileged clients given more access to instruct the server than the public clients or peers, the query is processed by the search interface or the peer to peer interface which make a request to the data access service to retrieve the required response. The data access service accesses the elastic search index which is bundled along with the installation of the server. The database response is converted to a JSON response and sent back to the client. If the client is a peer in the network requesting information, the data is sent to the peer instance for it to store in it's local database. For every client request, the new hashtags and text encountered are recursively added to the server queue for a given depth. Crawler applies the corresponding scrapers to convert HTML into necessary objects before adding them to the database.

A *campaign* is a targeted scraping method where the server is instructed to scrape a targeted number of messages from a given start date to an end date. A campaign has a theoretical infinite depth and a practical limitation to the size of the queue that can be maintained by the server. Any request to perform a campaign makes a call to the crawler and scraper respectively. Campaigns are great ways to initially collect data without having any focus on the data being collected. Tools like EVEtar [3] and TwitterEcho [1] are dedicated and focused scrapers which collect only specific type of data, the Loklak server instances can also perform such specific data collection but need more careful planning on the type of crawls performed from the server. In such situations, its not recommended to create a campaign unless the intention is to retrieve as much information as possible and then filter according to requirement.

## 3.3 Configurability and Extensibility

Modularity is at the core of the design of the server and thus can be easily extended to include other micro blogging website by implementing the corresponding scraper and adding it to the existing system. The generic queueing and database logic can be reused completely as is. Added to the ability to easily add a new scraper implementation to the system, there are existing modules integrated within the system to convert the response data into other data types like XML and modules which mimic the social networks look and feel[5]. It is also possible to include sentiment analysis and profanity analysis tools as a module to retrieve information like the language of the tweet, probability of profanity and sentiment of the tweet. The server contains a naive bayes classifier as a proof of concept to show emotion classification.

The system is easily configurable and scalable as it uses the same configuration architecture as elastic search which is extensively tested, reliable, secure and resilient. The fastest way to increase the storage size would be to just add another elastic search node to the cluster. Just like every other distributed system, horizontal scaling allows for larger amount of data to be stored whereas vertical scaling would result in higher redundancy but faster access to data and increases reliability in cases of disasters.

The system allows system administrators to maintain their own network topologies while running the system, keeping different factors like number of resources available, high reliability, data exchange cost and bandwidth costs in mind. This also allows administrators setting up these clusters to maintain them on multiple cloud services and integrate them to an on premise cluster for maximizing the amount of data that can be collected in a hybrid cloud manner.

## 3.4 Client Anonymity

The Loklak server allows open access to data as long as users are not trying to perform a DDoS attack on the network. Unlike the official twitter API which needs authentication tokens to retrieve data, the server instance allows users to directly use the API endpoints without any authentication to retrieve the data. It also does not record Client IP addresses when a search is done and at the same time does not log IP addresses from searchers.

---
[5]http://loklak.net/

In case you do not want to crawl and scrape the websites, the server has the ability to directly import existing data dumps from the peers thus creating a clone and not contributing to the network.

## 3.5 Production Architecture

The network topology for the production systems running the system is shaped like a tree where the central peer pushes data to the configured back end server. The central back end server remains purely as a storage system collecting tweets from the peers which do the scraping, crawling and data transfer work. Such a storage of open data is publicly available at the loklak website with the API service. There are four other peers which use the central peer as their main back end within the cluster and these peers are generally instructed by the central peer to perform intensive tasks of scraping and crawling. Periodically, the data is transferred from the local database of the peer to the central peer designated as the back end, the data is then flushed out of these peers in the cluster and the process continues thus having only one single storage for all the tweets.

Newer peers joining the network which spin up on various cloud services or hosted machines connect directly to the central peer as its main back end and contribute to the peer to peer network by crowd sourcing the data it collects to the central peer. The central server is a set of two physical server machines with a storage space of 4TB collective on both the servers with extensible storage ability. One instance of the server runs on each of the servers and the client requests are balanced to both using a load balancer. Each of the servers have their own elastic search index clusters which work as the database system. The database instances are connected to each other and balance the data storage through the nodes thus making it an extremely scalable system.

## 4. EVALUATION

In comparison to the existing crawlers the Loklak server behaves more effectively due to the crowd sourcing nature and the peer to peer architecture that it adapts from the design of bot nets. It could crawl 780 million tweets, 53 times more data than TwitterEcho [1] which collected a total of 14.5 million tweets in the same crawl time. Creating the API responses format to be exactly the same as that of the official twitter API has been a factor for open source communities using twitter streams to replace the endpoints with Loklak server endpoints. The ability of the server to be installed on all popular cloud services with just one singe click has contributed to many developers and enthusiasts running their own instances with ease and contributing to the main central back end. The elastic search stack has had a proven track record for being reliable and robust thus making it easy for other big data enthusiasts and communities to experiment and offer feedback.

The average time taken for the server to respond to a client query is 0.017752 seconds and can handle more than 300 simultaneous client and peer connections requesting and sending data. The system has been proven to be production ready and helps data scientists and analysts use Kibana's built in visualization capability to visualize the data.

## 5. CONCLUSION

In this paper we present the work on building and running distributed systems for scraping social network and micro blogging data. Building a peer to peer network of such servers and scrapers prevents the web application being scraped from blacklisting the instances. Loklak has the ability to scrape and store tweets while crowd sourcing them into one single database and allowing open access to it. The server has been successful in scraping and crawling Twitter and similar social networks and has given rise to a large community of developers and open data enthusiasts using such a system as a direct alternative to the twitter API. One year since inception, the peer to peer network has successfully collected 780 million tweets from twitter whereas 50 million tweets per day are sent out on Twitter[6] making it a total of 18250 million tweets in a year or 18.25 billion tweets a year. After one and half years of scraping and community collaboration to build the network, the Loklak server hit the 1 billion tweets mark and currently after two years hosts more than 1.45 billion tweets. This contributes to 0.0397% of the total volume of tweets that were posted on Twitter in 2 years but is a great repository of historical and current data for researchers and developers.

The system has been running for more than 2 years and contains a strong community of developers and open source contributors. We expect Loklak to be a go to tool for researchers and analysts to carry out research with Twitter and data from other such micro blogging services.

---

[6]https://blog.twitter.com/2011/numbers